\documentclass[sigconf, authorversion=true, nonacm]{acmart}

\AtBeginDocument{%
  }

\usepackage{multicol}
\usepackage{multirow}
\usepackage{nicematrix}
\usepackage{listings}
\usepackage[most]{tcolorbox}
\usepackage{float}
\usepackage{tikz}
\usetikzlibrary{shapes, arrows.meta, positioning, fit, backgrounds}

\newtcolorbox{gpdai-function}{
  colback=gray!10,
  colframe=gray!50,
  boxrule=0.25pt,
  arc=2mm,
  left=5pt,
  right=5pt,
  top=0pt,
  bottom=0pt,
  enhanced,
  before skip=5pt,
  after skip=5pt,
  fonttitle=\bfseries,
}

\colorlet{punct}{red!60!black}
\definecolor{background}{HTML}{EEEEEE}
\definecolor{delim}{RGB}{20,105,176}
\colorlet{numb}{magenta!60!black}

\lstdefinelanguage{json}{
    basicstyle=\normalfont\ttfamily,
    numbers=left,
    numberstyle=\scriptsize,
    stepnumber=1,
    numbersep=8pt,
    showstringspaces=false,
    breaklines=true,
    frame=lines,
    backgroundcolor=\color{background},
    literate=
     *{0}{{{\color{numb}0}}}{1}
      {1}{{{\color{numb}1}}}{1}
      {2}{{{\color{numb}2}}}{1}
      {3}{{{\color{numb}3}}}{1}
      {4}{{{\color{numb}4}}}{1}
      {5}{{{\color{numb}5}}}{1}
      {6}{{{\color{numb}6}}}{1}
      {7}{{{\color{numb}7}}}{1}
      {8}{{{\color{numb}8}}}{1}
      {9}{{{\color{numb}9}}}{1}
      {:}{{{\color{punct}{:}}}}{1}
      {,}{{{\color{punct}{,}}}}{1}
      {\{}{{{\color{delim}{\{}}}}{1}
      {\}}{{{\color{delim}{\}}}}}{1}
      {[}{{{\color{delim}{[}}}}{1}
      {]}{{{\color{delim}{]}}}}{1},
}

\lstdefinestyle{mypython}{
    language=Python,
    basicstyle=\normalfont\ttfamily,
    numbers=left,
    numberstyle=\scriptsize,
    stepnumber=1,
    numbersep=8pt,
    showstringspaces=false,
    breaklines=true,
    frame=lines,
    backgroundcolor=\color{background},
    keywordstyle=\color{blue}\bfseries,
    commentstyle=\color{gray}\itshape,
    stringstyle=\color{red},
    tabsize=2,
    showspaces=false,
    showstringspaces=false,
    captionpos=b
}

\newif\ifdraft
\drafttrue 

\ifdraft
  \usepackage[colorinlistoftodos, textwidth=3.2cm, loadshadowlibrary, shadow]{todonotes}
\else
  \usepackage[disable]{todonotes}
\fi

\ifdraft
  \newcommand{\gilles}[1]{\todo[inline,color=blue!40,size=\scriptsize]{Gilles: #1}}
  \newcommand{\mahmoud}[1]{\todo[inline,color=orange!40,size=\scriptsize]{Mahmoud: #1}}
  \newcommand{\gaspard}[1]{\todo[inline,color=green!40,size=\scriptsize]{Gaspard: #1}}
\else
  \newcommand{\gilles}[1]{}
  \newcommand{\mahmoud}[1]{}
  \newcommand{\gaspard}[1]{}
\fi

\def\gdfai{\texttt{GeoDataFrameAI}}
\def\gdf{\texttt{GeoDataFrame}}
\def\gdfs{\texttt{GeoDataFrames}}
\def\gpdai{\texttt{GeoPandas-AI}}
\def\gpd{\texttt{GeoPandas}}
\def\ai{AI Service}

\newcommand{\append}{\mathbin{\mathbin{\Vert}}}

\begin{document}

\title{GeoPandas-AI: A Smart Class Bringing LLM as Stateful AI Code Assistant}

\author{Gaspard Merten}
\affiliation{%
  \institution{Data Science and Engineering Lab, Université libre de Bruxelles}
  \city{Brussels}
  \country{Belgium}
}
\email{gaspard.merten@ulb.be}

\author{Gilles Dejaegere}
\affiliation{%
  \institution{Data Science and Engineering Lab, Université libre de Bruxelles}
  \city{Brussels}
  \country{Belgium}
}
\email{gilles.dejaegere@ulb.be}

\author{Mahmoud Sakr}
\affiliation{%
  \institution{Data Science and Engineering Lab, Université libre de Bruxelles}
  \city{Brussels}
  \country{Belgium}
}
\email{mahmoud.sakr@ulb.be}

\renewcommand{\shortauthors}{Merten et al.}

\begin{abstract}
Geospatial data analysis plays a crucial role in tackling intricate societal challenges such as urban planning and climate modeling.
However, employing tools like \gpd, a prominent Python library for geospatial data manipulation, necessitates expertise in complex domain-specific syntax and workflows.
\gpdai\ addresses this gap by integrating LLMs directly into the \gpd\ workflow, transforming the \gdf\ class into an intelligent, stateful class for both data analysis and geospatial code development.
This paper formalizes the design of such a smart class and provides an open-source implementation of GeoPandas-AI in PyPI package manager. 
Through its innovative combination of conversational interfaces and stateful exploitation of LLMs for code generation and data analysis, \gpdai\ introduces a new paradigm for code-copilots and instantiates it for geospatial development. 
\end{abstract}




\maketitle

\section{Introduction and System Overview}

Geospatial data analysis has become central to addressing complex societal challenges, from urban planning to climate modeling. However, leveraging tools like \gpd\ — a widely used Python library for geospatial data manipulation — requires mastering domain-specific syntax and workflows. 
\gpd\ integrates several Python libraries, acting as a core component in the Python geospatial ecosystem.
Its power for building geospatial solutions comes at the cost of a steep learning curve: developers must navigate several packages, interface the data across them, and learn intricate function parameters and syntax to build their data science pipelines. This complexity often forces practitioners to rely on fragmented documentation, coding blogs, or trial-and-error approaches, slowing down software development.

Recent advances in Large Language Models (LLMs) have demonstrated remarkable potential for code generation, enabling natural language queries to be translated into executable scripts. Tools like GitHub Copilot and ChatGPT excel at assisting developers by generating code or explaining logic. Further, many LLM models now natively support data analysis. One could upload their data and obtain summaries and insightful visualizations. 


\gpdai\ takes advantage of the LLM code generation capabilities by embedding it directly into the \gpd\ workflow. It extends the \gdf\ into a \emph{smart class}, capable of interacting with an LLM to generate executable code and data analysis tailored to user-defined tasks. \gpdai \ is intended to be used by Python programmers as a complement, not a replacement for, the popular geospatial data processing library \gpd. It makes \gpd\ DataFrame conversational, allowing developers to describe a Python function in natural language or ask data analytics questions about the \gdfs\ and get back the code to perform this task. For example, you can ask \gpdai\ to cluster the nearby features in a \gdf\ and filter out small clusters, and it will produce the code for this and return a \gdf\ containing the clustering results. 

\gpdai\ positions itself as an intelligent code co-pilot, yet it distinguishes from existing approaches as illustrated in Table~\ref{tab:landscape}. \gpdai\ occupies a novel position in the landscape of AI-assisted programming tools. In the top-left cell of the table, we find syntax-level code assistants such as GitHub Copilot. These tools provide support based on syntactic patterns and have access to the overall project code-base. Still, they do not consider runtime information, such as the actual schema or content of a GeoDataFrame. However, in geospatial programming, the choice and application of operations are tightly coupled with the structure and semantics of the data. For example, the set of applicable geometric operations varies significantly depending on whether the geometry column contains points or polygons. To compensate for this limitation, users must manually convey detailed knowledge about the GeoDataFrame’s schema and content to the assistant. This task becomes increasingly complex as the schema evolves or interacts with other variables, such as during join operations with other GeoDataFrames.
~
Further, some tools speak to the needs of data analysts, who lack programming skills but understand the data and the business domain well, represented in the bottom row of Table~\ref{tab:landscape}. Contemporary LLM-based chatbots (e.g., ChatGPT) can ingest data files and generate meaningful insights without requiring deep programming expertise. This has enabled such data analysts to perform exploratory data analysis through conversational prompts, often referred to as \emph{vibe coding} (bottom-right cell in the table). Furthermore, zero-code platforms like LIDA analyze the content of data sources and interactively suggest tailored analytical workflows to non-programming users. 
~
\gpdai\ bridges these two paradigms by supporting software developers while maintaining a tightly integrated understanding of code semantics, enabling developers to work more efficiently.

A typical user interaction consists of initializing a \linebreak \gdfai\ object, then an arbitrary long \emph{conversation} with it, and ending with code generation. This process is illustrated in Listing~\ref{lst:running_example}. 
In this example, the programmer instantiates a \linebreak \gdfai\ from an existing \gdf\ and performs a simple conversation to plot the data.
In the background, the \gdfai\ will communicate with the \ai\ to produce the desired code.
The programmer will instantiate the \linebreak\gdfai\ and call the first \texttt{chat()} function, yielding an initial code and result.
The programmer will then inspect the results produced and decide to indicate to the \ai\ to add a legend by using the \texttt{improve()} function.
Once satisfied with the result, the programmer calls \texttt{inject()}, which creates a new Python file including the member function \texttt{plot\_network()} with the last code instance, making it available for reuse and further manual edits.
It should also be understood that such a snippet of code is produced incrementally, i.e., after writing each \texttt{chat()} or \texttt{improve()} operation, the user checks the result before deciding whether further \texttt{improve()} operations are necessary. 
Once satisfied with the results, the user can continue with the \texttt{inject()} procedure.

\begin{lstlisting}[style=mypython, basicstyle=\ttfamily\small, caption={Running example for the use of \gpdai.}, label={lst:running_example}]
gdfai = gpdai.GeoDataFrameAI(stib_gdf, "GeoDataFrame for the network of public transport operator in Brussels.")
gdfai.chat("Plot the netword")
gdfai.improve("add a legend")
gdfai.inject("plot_network")
\end{lstlisting}

The novelty of \gpdai\ stems from its design, which integrates conversational interfaces directly encapsulated into the GeoDataFrame class. Unlike most AI code generators or assistants that operate externally within the development environment, \gpdai\ is executed natively in Python as part of the data structure itself. 
Further, \gpdai\ is a stateful assistant, allowing an incremental refinement of the generated code based on the history of past interactions.
~
To the best of the author's knowledge, this model for code generation and data analysis has not existed in the literature.


\begin{table}[htbp]
\centering
\begin{tabular}{|>{\raggedright\arraybackslash}p{1.4cm}|
                >{\raggedright\arraybackslash}p{3cm}|
                >{\raggedright\arraybackslash}p{3cm}|}
\hline
\textbf{User Type}        & \textbf{Syntax-Level Assistance} & \textbf{Semantic Understanding} \\
\hline
Software Developer        & GitHub Copilot, Cursor, Duet AI, Tabnine & \textbf{\gpdai} \\
\hline
Data Analyst              & Chat-GPT, Gemini, Claude, Mistral AI & LIDA, Tableau Einstein, ThoughtSpot, etc. \\
\hline
\end{tabular}
\caption{Landscape of AI code assistants.}
\label{tab:landscape}
\end{table}

Several research papers and studies have attempted to formalize or derive design principles for AI-powered code generators and copilots, especially focusing on human-AI interaction, developer experience, and tool usability. These principles are informed by empirical findings from user studies, field observations, and design evaluations of existing tools such as GitHub Copilot, Amazon CodeWhisperer, and other machine learning-based programming assistants. Table~\ref{tab:design_principles} summarizes the key design principles distilled from this body of work. 

\begin{table}[htbp]
\centering


\begin{tabular}{p{2cm} p{5.8cm}}
\hline
\textbf{Principle} & \textbf{Description} \\
\hline

Context Awareness~\cite{10.1145/3630106.3658984} & Must understand syntax, variable state, and context. \\

Low Cognitive Load~\cite{haque2025decodingdevelopercognitionage,10.1145/3630106.3658984} & Suggestions should be intuitive and require minimal mental effort. \\

Privacy~\cite{10.1007/978-3-031-66459-5_3} & Respect code ownership, and sensitive data handling. \\

Feedback Loops~\cite{Le2024,sun2025surveyneuralcodeintelligence} & Enable developers to accept, reject, or refine suggestions. \\

Seamless Integration~\cite{sun2025surveyneuralcodeintelligence} & Must fit naturally into the development flows. \\

\end{tabular}
\caption{Key Design Principles for AI-Powered Code Generators and Copilots}
\label{tab:design_principles}
\end{table}


\gpdai\ brings unique benefits with respect to the first three principles in Table~\ref{tab:design_principles}. 
~
For context awareness, \gpdai\ goes beyond basic syntax by grounding the LLM in automated metadata (such as column types and coordinate reference systems) and user-provided context. This contextual information is encoded in natural language descriptions (e.g., ``This dataset contains OpenStreetMap parks in Berlin with accessibility features'') and structured into domain-specific knowledge, such as the dataset’s intended use (e.g., urban mobility), constraints (e.g., ``exclude private parks''), and spatial properties. By providing this dual layer of context, \gpdai\ ensures that generated code adheres not only to geospatial best practices but also to the analyst’s specific intent.
~
This rich contextual grounding naturally supports the principle of low cognitive load. As shown in our use case examples in Section~\ref{sec:examples}, users often receive accurate and usable code on the first interaction. 
Furthermore, the system mitigates common LLM pitfalls such as hallucinated function names or invalid projections by validating outputs against the DataFrame’s schema and spatial constraints. For example, if a dataset uses a local coordinate system, \gpdai\ ensures distance calculations respect that CRS rather than defaulting to inaccurate latitude-longitude approximations.
~   
Regarding privacy, \gpdai\ offers a more secure approach than other tools in Table~\ref{tab:landscape}. Unlike AI copilots that may expose full source code to the LLM, \gpdai\ offers the possibility to the user to share, in addition to its prompt, only the DataFrame’s metadata and an eventual data sample. Compared to chatbots and general-purpose automated analytics tools, it gives full control regarding the data and metadata shared with the \ai. This selective exposure significantly reduces legal and privacy risks associated with AI-assisted coding.

However, two key design principles remain challenging in implementing \gpdai: feedback loops and seamless integration into development workflows. These lead to the following technical challenges:

    \noindent- \textbf{Supporting feedback loops} requires stateful interactions between the user and the \gpdai\ object. This demands a formal memory model to track conversation states and transitions, along with an intuitive API for users to manage and refine their interactions over time. 

    \noindent- \textbf{Seamless integration} into Python workflows presents another challenge. Since \gpdai\ is implemented as a class, its integration must occur through member functions. The design of these functions must balance minimalism with flexibility, supporting both script-based and notebook-style Python development environments while maintaining ease of use. 

    \noindent- \textbf{Ensuring consistency across runs} is complicated by the stochastic nature of LLMs. Even with the same input, outputs can vary depending on factors like temperature settings. Yet, users expect consistent results when rerunning prompts during a conversation till reaching the code generation. Maintaining reliable behavior across multiple executions is thus a critical concern. 

    \noindent- \textbf{Geospatial code generation} poses an additional challenge. While LLMs perform well on general programming tasks, they are less proficient in niche domains such as geospatial analysis. Although fine-tuned models and domain-specific foundation models have emerged in the geospatial community~\cite{hou2024geocode,zhang2024geogpt}, \gpdai\ focuses specifically on GeoPandas code generation and its ecosystem. To bridge this gap, we augment the LLM using domain-specific RAG and compiled curated examples tailored to GeoPandas workflows. 
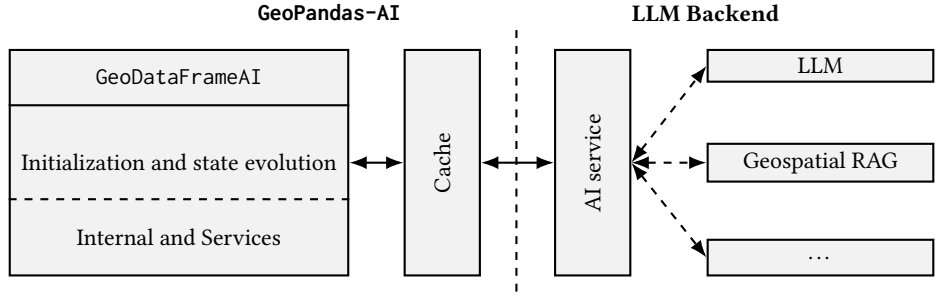
\begin{figure*}
\centering
\begin{tikzpicture}[
  service/.style={rectangle, minimum width=3.5cm, minimum height=0.8cm},
  component/.style={rectangle, draw, minimum width=3cm, minimum height=0.8cm, fill=gray!10},
  librarybox/.style={draw, 
                   very thick, 
                   minimum width=5cm, 
                   },
  classbox/.style={draw, 
                   thick, 
                   minimum width=4cm, 
                   minimum height=2.5cm
                   },
  dashedline/.style={draw, dashed, thick},
  arrow/.style={Latex-Latex, thick},
  dashedarrow/.style={Latex-Latex, thick, dashed},
  node distance=1cm and 1cm
]

  \node[label=above:] at (2, 0) {\textbf{\gpdai}};

\node[draw, 
      thick, 
      minimum width=4.5cm, 
      minimum height=3cm,
      fill=gray!10,
  ] (gdfai) at (0cm, -2cm) {};
  \node at ([yshift=-0.37cm, label=above]gdfai.north) {\gdfai}; 
  \draw[draw, thick] ([yshift=-0.75cm]gdfai.north west) -- ([yshift=-0.75cm]gdfai.north east);
  \node at ([yshift=-1.5cm, label=above]gdfai.north) {Initialization and state evolution}; 
  \draw[draw, thick, dashed] ([yshift=-2cm]gdfai.north west) -- ([yshift=-2cm]gdfai.north east);
  \node at ([yshift=-2.5cm, label=above]gdfai.north) {Internal and Services}; 

\node[draw, 
      thick, 
      minimum width=3cm,   
      minimum height=1cm,  
      fill=gray!10,
      rotate=90,
  ] (cache) at (3.5cm, -2cm) {Cache};

\draw[dashedline] ([yshift=0.25cm, xshift=1.5cm]cache.north east) -- ([yshift=-0.25cm, xshift=1.5cm]cache.north west);   

  \node[label=above:] at (7, 0) {\textbf{LLM Backend}};

\node[draw, 
      thick, 
      minimum width=3cm,   
      minimum height=1cm,  
      fill=gray!10,
      rotate=90,
  ] (backend) at (5.5cm, -2cm) {AI service};

\node[draw, 
      thick, 
      minimum width=3cm,   
      fill=gray!10,
      anchor=north west,
  ] (llm) at ([xshift=3cm]cache.south east) {LLM};

\node[draw, 
      thick, 
      minimum width=3cm,   
      fill=gray!10,
      anchor=west,
  ] (rag) at ([xshift=3cm]cache.south) {Geospatial RAG};

\node[draw, 
      thick, 
      minimum width=3cm,    
      minimum height=1.5em,    
      fill=gray!10,
      anchor=south west,
  ] (rest) at ([xshift=3cm]cache.south west) {\dots};

\draw[arrow] (gdfai.east) -- (cache.north);
\draw[arrow] (cache.south) -- (backend.north);
\draw[dashedarrow] (backend.south) -- (llm.west);
\draw[dashedarrow] (backend.south) -- (rag.west);
\draw[dashedarrow] (backend.south) -- (rest.west);

\end{tikzpicture}
\caption{Architecture representation of \gpdai.}
\label{fig:framework}
\end{figure*}

Accordingly, the main contributions in this paper are: (1) A design for a smart class mitigating the above challenges. Although out of the scope of this paper, one could consider this design as a template for developing other smart classes, e.g., for raster processing, for audio processing, etc, and (2) an open source implementation of the \gpdai\ class available in PyPi\footnote{\url{https://py pi.org/project/geopandas-ai/}}.

The \gpdai\ architecture, which addresses the above challenges, is illustrated in Figure~\ref{fig:framework}.
It consists of two main components: the \gpdai\ library and an LLM backend, also denoted AI service. 
The \gpdai\ library is the main component. 
It consists of a new \gdfai\ class and a caching system. 
The class is divided into two parts. 
The first part, denoted \emph{State and Interface}, is the component that the programmer will use and interact with. 
It defines the public functions and variables of \gdfai\ class.
The other part of the class, denoted \emph{Internal and Services}, contains the different components and logic not exposed to the programmer. 
These are necessary to communicate with the \ai\ and create the desired code.
The second component of the ecosystem is its LLM backend, also denoted the \ai. This \ai\ can be any system capable of producing code from a prompt. This definition includes RAG-powered LLMs, fine-tuned LLMs, etc. One key responsibility of the \ai\ is to deal with discrepancies to minimize the effect of hallucination, inconsistency, wrong answers, etc. 



The rest of this paper is organized as follows.
\gdfai\ is detailed in Sections \ref{sec:geopandasai_environment} and \ref{sec:internals}, with Section \ref{sec:geopandasai_environment} formally presenting its exposed interface and its state evolution model while Section \ref{sec:internals} presents how the \gdfai\ class functions internally: how it communicates with and uses the \ai\ to generate the desired code.
Section~\ref{sec:caching} motivates the need for caching and how it is implemented in the \gpdai\ library.
The \ai\ and how it is specialized in generating geospatial Python code are detailed in Section \ref{sec:backend}.
Then, the main use case scenarios are presented in Section~\ref{sec:use_cases}.
Some examples of practical usage of \gpdai\ are provided in Section~\ref{sec:examples}.
Finally, Section~\ref{sec:ccl} concludes this work and provides avenues of future research.

\section{Related Works}


LLMs have also proven capable of generating high-quality source code. Further, specialized LLM models optimized for code generation tasks have emerged, such as CodeGemma\cite{codegemmateam2024codegemmaopencodemodels} from Google and Codstral\footnote{\url{https://mistral.ai/news/codestral}} from Mistral AI. A comprehensive survey by Jiang et al. examined numerous Code LLMs, which are models specifically fine-tuned for translating natural language into source code \cite{jiang2024surveylargelanguagemodels}. 
Recent studies further benchmark these models across various tasks and programming languages \cite{yan2024codescopeexecutionbasedmultilingualmultitask, li2024evocodebenchevolvingcodegeneration}.

Although web-based conversational interfaces like \texttt{ChatGPT.com} are widely used by developers \cite{khojah2024beyond}, they lack the deep integration expected from modern IDEs. This gap has led to the rise of syntax-level assistants embedded within development environments. As summarized in Table~\ref{tab:landscape}, such tools provide context-aware code completion and suggestions based on static analysis of the codebase. These tools increasingly adopt agent-like behaviors \cite{yao2023react}, enabling them to perform complex multistep tasks with minimal human intervention. However, a key limitation remains: these tools operate outside the runtime environment and are thus unaware of the runtime context.

Numerous tools have been specifically designed to support data scientists by leveraging large language models (LLMs) for tasks such as data exploration and visualization. For example, Wang et al. proposed an interactive online visualization platform that employs LLMs to streamline data analysis workflows \cite{wang2024dataformulator2iteratively}. These tools span both academic research systems, such as LIDA \cite{dibia2023lida}, and commercial solutions like Tableau Einstein and ThoughtSpot (see Table~\ref{tab:landscape}).

Geospatial data requires specialized domain knowledge.
As such, specialized LLMs have been proposed.
For instance, Li et al. proposed Geo-Llama \cite{li2024geo}, a fine-tuned LLM for the generation of trajectories.
In \cite{xue2025transforming}, Xue et al. propose an overview of recent efforts and challenges in the exploitation of LLMs for smarter mobility systems.
These, however, are not designed to produce code.
To solve this problem, researchers have also proposed both dedicated LLMs and integrated tools.
GeoCode-GPT \cite{hou2024geocode} is an LLM designed specifically for geospatial code generation.
On the tooling side, GeoGPT \cite{zhang2024geogpt} provides a framework that allows users to interact with various GIS tools through a single natural language interface powered by LLMs. 
While these two LLM models have been trained/fine-tuned on the specifics of Geospatial, the part related to the \gpd \ ecosystem is limited.



\section{Initialization and State Evolution of \gdfai\ Objects} \label{sec:geopandasai_environment}

A key challenge in inserting \gpdai\ into the geospatial developer's programming workflow is to make its front-end user-friendly and well-integrated.
To address this, we introduce a minimal extension of the \texttt{GeoDataFrame} class, \gdfai.
This \gdfai\ smart class is the entry point for \gpdai\, enabling developers to interact with an \ai\ through natural language.

Objects of this \gdfai\ class are defined by a state, denoted $\mathcal{S}$, which is defined as a tuple: 
\begin{equation}
	\mathcal{S} = \langle G, \mathcal{M}_G, \mathcal{H}, \mathcal{T}, \mathcal{R} \rangle
  \label{eq:state}
\end{equation}
where:
\begin{itemize}
  \item $G$: A classical input GeoDataFrame. 
  \item $\mathcal{M}_G$: Metadata derived from $G$ both automatically (e.g., \linebreak schema, summary statistics, spatial reference), from a user input description of $G$ as well as potentially from other \gdfai\ objects. 
	\item $\mathcal{H}$: History of prior interactions, i.e., user queries and generated code.
	\item $\mathcal{T}$: \textit{Tool State} — A set of permissible Python packages that the model is allowed to invoke in the generated code. Initially 
    $\mathcal{T}= \mathcal{T}^0= \{$contextily, pandas, matplotlib, folium, geopandas$\}$.

	\item $\mathcal{R}$: \textit{Return-Type Set} — A set of possible response types. Initially  
	      $\mathcal{R} = \mathcal{R}^0  = \{$int,
            float,
            str,
            bool,
            list,
            dict,
            geopandas.GeoDataFrame,
            pandas.DataFrame,
            folium.Map,\linebreak
            matplotlib.Figure$\}$.
\end{itemize}

In the rest of this text, we will denote $G, \mathcal{M}_G, \mathcal{H}, \mathcal{T}, \mathcal{R}$ as variables. 
These can be split into three categories.
$G$ and $\mathcal{M}_G$ enable the environment to maintain awareness of the nature of the dataset and its structure, while $\mathcal{H}$ allows the environment to maintain awareness of the user’s workflow history.
Finally, $\mathcal{T}$ and $\mathcal{R}$ restrict how the \ai\ can operate.

The remainder of this section describes the initialization of \gdfai\ and the state evolution during user interactions, while Section \ref{sec:internals} presents how the class functions internally.

\subsection{Initialization of a \gdfai\ Object} \label{sec:initialization}

To create an initial \gdfai\ object, a constructor can be called with a \gdf\ $G$ and an optional user-provided natural language description $d$. 
This constructor is defined as:
\begin{gpdai-function}
  \begin{equation}
    \gdfai(G, d) \rightarrow  \mathcal{S}^{0} = \langle G, \mathcal{M}_G^{0}, \emptyset, \mathcal{T}^0, \mathcal{R}^0\rangle
    \label{eq:constructor}
  \end{equation}
\end{gpdai-function}
This is illustrated on the first line of the running example (Listing~\ref{lst:running_example}).

At this stage: 
\begin{equation}
  \mathcal{M}_G^0 = (A_{G}, d, \mathcal{L}^0) = (A_{G}, d, \emptyset)
	\label{eq:M_G}
\end{equation}
with $ A_{G} $ and $\mathcal{L}^0$ representing automated metadata respectively extracted from ${G}$, and to be extracted from other \gdfai\ objects (see Section~\ref{sec:chat}). 
~
The automated metadata $ \mathcal{A}_G $ is derived directly from $ G $ without user intervention. It includes:

\noindent- \textbf{Schema}: Column names, data types (e.g., string, numeric), and constraints (e.g., ``column `population' is non-negative'').

\noindent- \textbf{Statistical Summaries}: For numeric columns, this includes measures such as mean, variance, minimum, and maximum values.

\noindent- \textbf{Spatial Properties}: For the geometry attribute, this captures geometric and coordinate information, such as the Coordinate Reference System (CRS), the geometry type, the spatial extent of the dataset, etc.

Valorizing the LLM's natural language understanding capabilities, $ d $ is kept unprocessed and unstructured --- in natural language as given by the user.
For instance, a user might provide the description:  

  \texttt{``This dataset shows all parks in Brussels, sourced from OpenStreetMap, and includes details like their names, sizes, and accessibility features''}.
Nevertheless, it distills the input into three broad themes to guide the model behavior: \\
\noindent- \textbf{The domain} refers to the field or topic the data belongs to. If the description mentions ``parks in Brussels'' the AI recognizes this as urban planning or environmental science. This helps it prioritize, for instance, visualization styles relevant to that area. \\
\noindent- \textbf{The purpose}  is what the dataset is meant to achieve. For instance, if the purpose is real-time monitoring, the AI uses this to align its responses with goals, such as dynamic visualizations or live data updates. \\
\noindent- \textbf{The constraints} or rules inferred from your description. 
  If the user notes that data comes from ``OpenStreetMap'' or ``only includes parks larger than 1 acre,'' the AI understands these boundaries. This prevents errors, such as suggesting tools incompatible with open-source datasets or ignoring size filters.

While $\mathcal{L}^0$ is still an empty set at this stage, the variable $\mathcal{L}$ is designed to contain a list of other \gdfai\ objects to be used by the \ai. 
This is useful whenever the programmer wants to analyze multiple DataFrames or perform a join operation.

We will denote $\mathcal{S}^0$ as the initial state.

\subsection{State Evolution Trough AI Assisted Development} \label{sec:chat}

\begin{figure}
    \centering
    \includegraphics[width=0.85\linewidth]{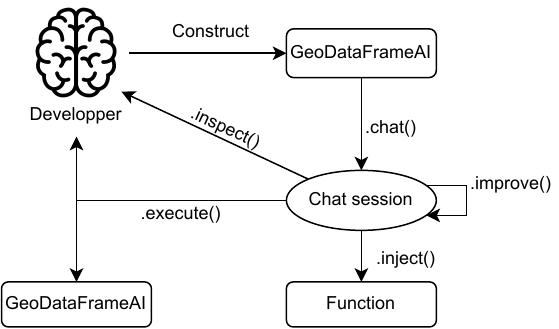}
    \caption{Possible \gpdai\ development work flows.}
    \label{fig:user-flow}
\end{figure}

In order to converse with the \gdfai\ object, we design a set of functions. 
The possible workflows in a chatting session are illustrated in Figure~\ref{fig:user-flow}.
Once a \gdfai\ object is constructed, two functions allow the invocation of the \ai: \texttt{chat()} and \texttt{improve()}.
A call to either of these functions results in the \gdfai\ interacting with the \ai\ and generating some code. 
This code can then be:
\begin{itemize}
  \item displayed to the programmer with the \texttt{inspect()} function. 
  \item executed with the \texttt{execute()} function. The result, which could be a new \gdfai\ object, a leaflet map, etc, is then returned to the developer.
  \item saved into a new Python function with the \linebreak \texttt{inject.()} function.
\end{itemize}
The default behavior is to always execute the generated code. This behavior can, however, be overwritten as shown in Section \ref{sec:security}. 
~
The functions are formally defined hereunder:

\paragraph{\texttt{chat()}} The \texttt{chat()} function is used to initialize or reinitialize a conversation. 
The \texttt{chat()} method of an \gdfai\ object must receive a user query, denoted $q^1$, representing a request. 
An example is provided in the second row of the running example (Listing~\ref{lst:running_example}), where $q^1$ is ``Plot the network''.
The \texttt{chat()} function also accepts some optional parameters. 
These are a tool set, a return type set as well as a list of other \gdfai s, respectively denoted $\mathcal{T}$, $\mathcal{R}$ and $\mathcal{L}$.   
By indicating the optional parameters in between square brackets, the \texttt{chat()} function is formally defined by:
\begin{gpdai-function}
\begin{equation}
  \mathcal{S}^i\texttt{.chat}(q^1, [\mathcal{T}, \mathcal{R}, \mathcal{L}]) \rightarrow \mathcal{S}^1
	\label{eq:chat}
\end{equation}
\end{gpdai-function}
The index $i$ being an integer greater than or equal to zero, as \texttt{chat()} can be called on the initial state or any other state.
~
The transformation is applied in multiple steps as follows:
\begin{enumerate}
  \item The state is reset to $S^0$.
  \item If $\mathcal{T}$ is provided then $\mathcal{T}^1$ is set to $\mathcal{T}$, otherwise to $\mathcal{T}^0$. Similarly for $\mathcal{R}^1$ and $\mathcal{L}^1$.
  \item $\mathcal{M}_G^1$ is set to $(A_{G}, d, \mathcal{L}^1)$.
  \item $G$, $M_G^1$, $\mathcal{T}^1$, $\mathcal{R}^1$ and $q^1$ are used by the \ai\ to generate the code $c^1$. 
  Details about the process used to generate the code can be found in Section~\ref{sec:code_generation}.
  \item The history of interaction $\mathcal{H}^1$ is set to $\langle \{q^1, c^1\}\rangle$. 
\end{enumerate}

After these successive transformations, the state $S^1$ is entirely defined by:
\begin{equation}
  \mathcal{S}^1 = \langle G, \mathcal{M}_G^1, \mathcal{H}^1, \mathcal{T}^1, \mathcal{R}^1\rangle
	\label{eq:chat_update_state}
\end{equation}

The default behavior of \texttt{chat()} is to automatically execute the last generated code, i.e., by calling the \texttt{execute()} function.

\paragraph{\texttt{execute()}}The \texttt{execute()} function is defined as follows:
\begin{gpdai-function}
\begin{equation}
  \mathcal{S}^i\texttt{.execute}() \rightarrow o^i
	\label{eq:execute}
\end{equation}
\end{gpdai-function}
\noindent with $i$ being a strictly positive integer, which is equal to $1$ when the function is called directly after the \texttt{chat()} function. The function returns $o^i$, which is the result of the execution of the code $c^i$ on the GeoDataFrame $G$ (using eventually the other GeoDataFrames provided in $\mathcal{L}^i$).
In generating $c^i$, the AI is instructed that the type of $o^i$ must be contained in $\mathcal{R}^i$.
In cases where $|\mathcal{R}^i| > 1$, the exact return type of $o^i$ is determined by the \ai\ using the queries present in $\mathcal{H}^i$.
If this type is the \gdf\ class, then $o^i$ will be a newly generated \gdfai\ object in an initial state:
\begin{equation}
  o^i = \mathcal{S}^{\prime0} = \langle G^\prime, \mathcal{M}_{G^\prime}^0, \mathcal{H}^0, \mathcal{T}^0, \mathcal{R}^0\rangle
  \label{eq:execution_output_gdfai}
\end{equation}
With $G^\prime$ being the \gdf\ generated by the code $c^i$.
It is important to note that $o^i = \mathcal{S}^{\prime0}$ is a new object. It is potentially a new \gdfai\ based on a modified DataFrame ($G^\prime$), however, the state $\mathcal{S}^i$ is not modified by the execution.

\paragraph{\texttt{inspect()}} The \texttt{inspect()} function allows users to check the generated code history at any time in the conversation. It is thus defined as follows:
\begin{gpdai-function}
\begin{equation}
  \mathcal{S}^i\texttt{.inspect}() \rightarrow \mathcal{H}^i
	\label{eq:inspect}
\end{equation}
\end{gpdai-function}

\paragraph{\texttt{improve()}} 
As long as the user is not satisfied with the results (either by inspecting or executing the code), they may refine the output proposed by the \ai. 
The \texttt{improve()} function must receive a user query $q^i$. 
This user query should not contain a new request (such as for the \texttt{chat()} function) but rather some indication guiding the \ai\ into producing better code. 
An example is provided in the second row of the running example (Listing~\ref{lst:running_example}), where $q^i$ (in this case $q^2$) requires the \ai\ to add a legend to its previously generated plot.
With this function, the user seeks to guide the model towards adjusting the generated code to their specific intent.
Such as \texttt{chat()}, \texttt{improve()} also accepts a tool set, a return type set, or a list of other \gdfai s as optional parameters:
\begin{gpdai-function}
\begin{equation}
  \mathcal{S}^{i-1}\texttt{.improve}(q^{i}, [\mathcal{T}, \mathcal{R}, \mathcal{L}]) \rightarrow \mathcal{S}^i
	\label{eq:improve}
\end{equation}
\end{gpdai-function}
\noindent with $i$ being the strictly positive index corresponding to the iteration of the \texttt{improve()} call.

The \texttt{improve()} function works similarly to \texttt{chat()}, with the difference that the state is not reset to the initial state:
\begin{enumerate}
  \item If $\mathcal{T}$ is provided then $\mathcal{T}^i$ is set to $\mathcal{T}$, otherwise to $\mathcal{T}^{i-1}$. Similarly for $\mathcal{R}^i$ and $\mathcal{L}^i$.
  \item $\mathcal{M}_G^i$ is set to $(A_{G}, d, \mathcal{L}^i)$.
  \item $G$, $M_G^i$, $\mathcal{T}^i$, $\mathcal{R}^i$ and $q^i$ are used by the \ai \ to generate the code $c^i$. 
  \item $\langle \{q^i, c^i\}\rangle$ are append to the history of interactions: 
  \begin{equation}
    \mathcal{H}^i = \mathcal{H}^{i-1} \append \{q^i, c^i\} 
    \label{eq:improve_append_history}
  \end{equation}
\end{enumerate}

Similar to the \texttt{chat()} function, the default behavior of \linebreak \gpdai\ is to automatically invoke \texttt{execute()} after calling the \texttt{improve()} function.

\paragraph{\texttt{inject()}} Once the programmer is satisfied with the results, they can call the \texttt{inject()} function. 
This function writes the last generated code as a function in a local Python script, allowing further usage on different datasets.
It takes a string parameter representing the desired name for the generated function:
\begin{gpdai-function}
\begin{equation}
  \mathcal{S}^{i}\texttt{.inject}(\texttt{function\_name}) 
	\label{eq:inject}
\end{equation}
\end{gpdai-function}

%

\section{Internal and Services} \label{sec:internals}

This section presents how the \gdfai\ class operates internally, i.e., how \gdfai\ communicates with the \ai\ to generate code.
These are a set of private functions in \linebreak\gpdai\ that are not exposed to the user.

\subsection{Templating} \label{sec:template}
To interact with an LLM, it is necessary to transform the state $\mathcal{S}^i$ and arguments of both \texttt{chat()} and \texttt{improve()} into natural language.  
This is done using a lightweight templating format inspired by well-established work\footnote{Django \url{https://docs.djangoproject.com/en/5.2/ref/templates/language/}, Jinja2 \url{https://jinja.palletsprojects.com/}}. 
The template language is designed based on the JSON format and adheres to a fixed schema designed for simplicity and clarity. Each template consists of an ordered list of message objects, defined as follows:

\begin{lstlisting}[language=json,firstnumber=1, basicstyle=\ttfamily\small]
{"messages": [{"role": "system" | "user", 
               "content": string}]}
\end{lstlisting}

Every object within the \texttt{messages} array contains two fields: \texttt{role}, indicating the speaker (either \texttt{"system"} or \texttt{"user"}), and \texttt{content}, a string representing the message text. The \texttt{content} string may also include specific patterns that indicate template variables to be injected during compilation.
A simple example of a template with variables can be seen hereunder:
\begin{lstlisting}[language=json,firstnumber=1, basicstyle=\ttfamily\small]
{"messages": [
    {"role": "system",
     "content": "You are a helpful coding assistant, produce a code answering the user prompt"},
    { "role": "user",
      "content": "Please generate a function def 'execute(args)' answering: {{prompt}}"}]}
\end{lstlisting}

\subsection{Code Generation} \label{sec:code_generation}

\gpdai\ relies on an \ai\ to generate executable Python code from natural language instructions. This section details the process by which code is synthesized, validated, and optionally refined across the two core interfaces: \texttt{chat()} and \texttt{improve()}.
The main steps to generate a code $c^i$ based on the state $S^{i-1}$ and the query $q^i$ are explained next.

\paragraph{1. Resolution of the return type.}  
When multiple return types are permitted (i.e., $|\mathcal{R}^{i}| > 1$), a dedicated \texttt{determine\_type()} function is executed. 
This function queries the \ai\ to define the most appropriate return type. 
This query generates a prompt using its associated template filled with $\mathcal{R}^{i}$ and $q^i$ and sends it to the \ai.
The output is then parsed using a regex to obtain the desired output type (denoted $r^i$).

\paragraph{2. Generation of the code-generation prompt.}  
To compile the prompt used to generate $c^i$, the templates for the \texttt{chat()} and \texttt{improve()} functions are filled with:
\begin{itemize}
    \item $r^i$: the return type obtained at step 1;
    \item $\mathcal{M}^i$: the metadata about the \gdf s (e.g., column types, CRS);
    \item $\mathcal{T}^i$: the currently allowed toolsets or libraries;
    \item $\mathcal{R}^i$: the currently expected return type;
    \item $\mathcal{H}^i$: the history of interactions (only for \texttt{improve()}) 
\end{itemize}
The details about how this information is encoded in natural languages are skipped here but can be found in the git repository of the project\footnote{\url{https://github.com/GaspardMerten/geopandas-ai}}.  
However, it should be mentioned that both templates also indicate to the LLM to generate a function of the form:
\begin{equation}
  \texttt{def execute(df1, [df2, \dots])} \rightarrow r^i
  \label{eq:funct_signature}
\end{equation}
With \texttt{df1} representing the \gdf\ $G$ while \texttt{[df2, \dots]} represent the other potentially linked \gdfai s contained in $\mathcal{L}^i$ (see Section \ref{sec:chat}).
This ensures that the generated code conforms to a known signature, which is validated in subsequent steps.

\paragraph{3. Execution and error-driven retries.}  
Once the prompt has been generated by filling in the template, it can be submitted to the LLM.
A regular expression is used to extract the code snippet $c^{i}$ from the generated answer.
This also implicitly ensures that the $c^i$ adheres to the requested function signature. This step serves to both constrain and verify the \ai{}'s output.
The code is then executed either against synthetic data derived from the original \gdf\ $G$, or a small excerpt of $G$ (see Section~\ref{sec:security}).
This serves two goals:
 
\begin{enumerate}
    \item Verify that the function runs without exceptions;
    \item Ensure the result matches the expected return type;
\end{enumerate}

If execution fails due to syntax errors, exceptions, or invalid return types, the system issues a new prompt using a retry template. This prompt includes the original code and the associated traceback, asking the \ai \ to revise its response. The conversation history is augmented with each retry attempt, enabling successive refinement. This retry loop is attempted up to five times. If no valid result is obtained, the system raises an error containing the last code produced and halts the process.

\subsection{Secure Communication Between \gdfai\ and the \ai} \label{sec:security}
Since the smart \gdfai\ class must execute code generated by AI services, it introduces several risks. These originate from the nature of such services: many operate in remote, opaque cloud environments, meaning that any data sent to them may be exposed to confidentiality breaches \cite{risksAiGeneratedCode}. Even when AI runs locally, the code it produces may be unsafe. One study found 40\% of GitHub Copilot-generated code to be vulnerable \cite{aSleepAtTheKeyboard}. Malicious actors can also perform data poisoning, leading to a compromised training process and potentially harmful outputs \cite{dataPoisoning}. Moreover, developers often lack visibility into what the AI is doing, making it difficult to trust and audit the interaction.

These issues impose three requirements for the front-end environment to mitigate these risks.

\paragraph{Data privacy.}
The content of the \gdf\ should not be sent directly to the AI service. This is to mitigate the confidentiality of the data from a legal perspective\footnote{\url{https://gdpr-info.eu/}} but also to prevent another type of data leakage.

\paragraph{Source code privacy}
In contrast with systems such as GitHub Copilot, the source code in which \gpdai\ is involved should not, in any case, be transmitted to the AI service. There are several reasons to keep source code private: (1) it may contain secret keys, such as API credentials\footnote{For example, an API key to an online service}; (2) it may include intellectual property; (3) it may embed trade secrets; and (4) it may play a role in security through obfuscation.

\paragraph{Execute code isolation}
AI-generated code, whether due to hallucination or intentional manipulation, can pose risks. It may modify critical system files, execute harmful operations, or compromise the host operating system. Additionally, it can be exploited to bypass data protection mechanisms—for example, by issuing unauthorized HTTP requests to exfiltrate sensitive information. These risks highlight the need for strict code execution isolation, where AI-generated code runs in a sandboxed environment without access to the host OS or networks.\\

To address these three points, \gpdai\ implements the following communication framework with the \ai.
The developer can instantiate a \gdfai\ from the \gdf\ $G$ anywhere in their code. At this point, no code or request has been executed. 
Once the programmer starts a \texttt{chat()} interaction, the system will need to generate a unified textual description (UTD) of the \gdfai. 
By default, an object of the class \texttt{PublicDescriptor} handles the generation of the UTD from the potential user-provided description, the metadata, as well as an excerpt of $G$. 
Users handling sensitive schema, which should not be sent to the \ai, can build their own custom \texttt{Descriptor} class from the base abstract class \texttt{ADescriptor}.
An object of this custom-made class can then be passed to the configuration of \gpdai\, which will then be used to generate the UTD.
The user can also include, in this descriptor, artificial data generators to simulate the \gdf\ schema and content.
This provides the user with the entire flexibility of determining what information should be sent to the \ai\ from the \gdf. 
Once the UTD is transmitted to the \ai\, the latter will generate an answer from which the code will be extracted. 
Executing it inline may lead to data exposure.
Ideally, \gpdai\ could serialize the DataFrames and pass them to a fully isolated Docker or Firecracker instance, reset after each run. 
The results should then also be serialized back. 
This would, however, complicate the installation process and execution of \gpdai.
Instead, two solutions are proposed to mitigate the execution of harmful Python code. 
The first one is introduced by the abstraction of the \ai\ to \gpdai. 
This means that the user may use as an \ai\ service any trusted LLM, potentially run locally.
The second one consists of a \texttt{safe\_mode} in which no code is executed automatically.
This allows the user to manually inspect generated code before executing it.

These steps allow for (1) ensuring data privacy, (2) preventing code-scope access, and (3) mitigating the execution of malicious code, hence solving issues currently faced by AI services.

\section{Caching} \label{sec:caching}

\gpdai\ relies on Large Language Models (LLMs) for dynamic code generation. While this provides the necessary creativity and problem-solving capabilities, it also inherits several limitations of LLM-based systems:

\noindent-  \textbf{Variability}: LLMs use stochastic sampling techniques, typically governed by a \emph{temperature} parameter, to balance creativity and correctness. This randomness can lead to inconsistent outputs across identical prompts. In the context of \gpdai, this undermines reproducibility, especially when using chained calls like \texttt{chat()} followed by multiple \texttt{improve()} steps, which implicitly rely on deterministic intermediate outputs.

\noindent-  \textbf{Latency}: Cloud-based or local LLMs often introduce noticeable response delays, which can accumulate across multiple queries, degrading the interactive experience and lengthening execution time.

\noindent-  \textbf{Cost}: Each LLM query incurs computational and sometimes financial cost. As the metadata $\mathcal{M}$ and history $\mathcal{H}$ grow, so do the token lengths and associated inference time. This increases usage cost and contributes to the environmental footprint of repeated LLM queries.

To address these challenges, we introduce a caching mechanism bringing determinism to \gpdai\ and reducing redundant LLM inferences, which, in fine, lowers the latency and cost of \gpdai.

\subsection{Caching Mechanism} \label{sec:cache}

Initially, when a user calls \texttt{chat()} or \texttt{improve()}, it triggers a transition from state $S^{i-1}$ to $S^i$ based on the instructions provided by the user. The main driver of this transition is the generation of $c^i$ by the \ai. 

This generation step is the primary source of variability, latency, and cost. Hence, the proposed caching mechanism targets this part of the transition to improve efficiency.

We define the caching mechanism as the two usual functions $set(key, value)$ and $get(key) \rightarrow value$, respectively updating and retrieving the value (in our case $c^i$) from a persistent memory unit (e.g, file-system storage). To adapt this caching mechanism to \gpdai, we must store both the previous state $S^{i-1}$, as well as the various arguments, leading to the generation of $c^{i}$, which eventually allows transitioning to $S^{i}$. This is done using the function \texttt{buildStateKey}:
\begin{gpdai-function}
\begin{equation}
  buildStateKey(S^{i-1}, q^{i}, \mathcal{T}^{i}, \mathcal{R}^{i}, c^{i}) \rightarrow k^i
\end{equation}
\end{gpdai-function}

The arguments are:
\begin{enumerate}
  \item $S^{i-1}$: the previous state of the \gdfai\ containing the entire history, including the previous sequence of queries and generated codes; $S^{i-1} \supseteq \{ q^j, c^j \mid 1 \leq j \le i \}$\footnote{We only use a subset of $S^i$ in \texttt{buildStateKey}, excluding $G$, as it is already represented through $\mathcal{M}^0_G$}.
    \item $q^{i}$: the prompt provided by the user to transition from $S^{i-1}$ (or $S^0$ in the case of a \texttt{chat()} call) to $ S^i$. 
    \item $\mathcal{R}^{i}$: the desired possible return types for $c^{i}$.
    \item $\mathcal{T}^{i}$: the allowed tool set for $c^{i}$.
\end{enumerate}

\paragraph{Cache Consistency.}
Because each key embeds the full sequence of prior queries and responses, which are contained in $S^{i-1}$, as well as the user instructions to go from $S^{i-1}$ to $S^i$, any modification to an earlier prompt (e.g., changing $q^{i-2}$) yields a different key, ensuring cache consistency. This guarantees semantic consistency and avoids unintended reuse of outdated outputs.

\paragraph{Usage.}
During execution of \texttt{chat()} or \texttt{improve()}, \linebreak \gpdai\ computes the corresponding key $k^i$. 
If $get(k^i)$ exists, the cached output $c^i$ is returned directly. 
Otherwise, the \ai\ is invoked to produce $c^i$, and the result is stored via \texttt{set($k^i$, $c^i$)} for future reuse.


\paragraph{Limitations.}
The primary limitation of the current caching mechanism is the absence of an automatic or periodic cache invalidation policy, which could theoretically lead to unbounded persistent storage growth. However, this risk is mitigated by the typically small size of each cache entry $c^i$, which is expected not to exceed a few dozen lines of code. A practical upper bound can be estimated from the maximum generation capacity of the underlying model. For example, Gemini 2.0 models are limited to 8,192 tokens per output. Given that one token corresponds to approximately four characters, this yields an upper limit of 32,768 characters, or roughly 32 KB in UTF-8 encoding. As such, the storage footprint remains modest in typical use cases. Nevertheless, to manage caching explicitly, we provide two tools:
\begin{enumerate}
    \item An instance method \texttt{reset\_cache(chat\_wise: boolean)} that clears the cache for a specific \gdfai\ instance. The \texttt{chat\_wise} parameter determines whether to delete only the current chat's cache or all chats associated with that instance.
    \item A global function \texttt{reset\_cache()} that purges all \linebreak \gpdai\ caches across the current environment.
\end{enumerate}

\section{LLM Backend} \label{sec:backend}

As outlined in the introduction, \gpdai\ exploits a default \ai{} which has been further specialized for the generation of GeoPandas Python code. 
The \gpdai\ library, however, is backend-agnostic and may exploit any user-provided \ai\ with various forms ranging from remote API calls to fully local language models. 
This flexibility is achieved through the use of the \texttt{litellm} Python package\footnote{\url{https://www.litellm.ai/}}, which serves as an abstraction layer over a wide array of LLM providers and configurations. 
As a result, \gpdai\ decouples its functionality from any specific vendor or model and supports diverse deployment scenarios.

While general-purpose LLMs are increasingly proficient at generating code, they often lack the domain expertise required to reason about geospatial concepts such as projections, topology, or geometry validity.
They may also struggle to use specialized tools like \gpd, Shapely, or Pyproj effectively.
To evaluate the impact of backend design on the system’s performance and geospatial reliability, we experimented with three distinct configurations.

\paragraph{Raw LLM}  
Our baseline configuration leveraged the Gemini 2.0 Flash model\footnote{\url{https://cloud.google.com/vertex-ai/generative-ai/docs/models/gemini/2-0-flash}} via Google Cloud’s Vertex AI service. This setup demonstrated that general-purpose LLMs can produce syntactically valid code. However, it frequently failed to capture the user's intent and often required multiple \texttt{improve()} iterations to converge to an acceptable solution. Additionally, geospatial subtleties such as CRS mismatches or geometry simplification were inconsistently addressed.

\paragraph{Fine-Tuned LLM}  
To enhance geospatial understanding, we fine-tuned the same Gemini model on a subset dataset of Stack Overflow answers\cite{lozhkov2024starcoder}. 
This subset is constructed by filtering answers containing ``geopandas'' in Python.
While the model preserved its ability to generate correct code, it did not significantly outperform the base model in terms of accuracy or concordance with user queries.
Moreover, the fine-tuning process introduced a regression: the model lost its ability to handle non-code prompts (e.g., type disambiguation in \texttt{determine\_type()}) due to overfitting on code-generation patterns.
This necessitated maintaining two separate models: one for logic generation, another for other instructions, which increased system complexity.

\paragraph{RAG-enhanced LLM}  
In a third and more promising approach, we implemented a Retrieval-Augmented Generation (RAG) backend. This system indexed a high-quality corpus of geospatial programming examples sourced from permissively licensed datasets and documentation. 
At runtime, semantically similar examples were retrieved and embedded alongside the user query in the prompt sent to the LLM. This configuration markedly improved the system's performance: it reduced the number of \texttt{improve()} calls needed, aligned better with domain best practices, and led to more robust, intention-aligned code generation. 

\section{Usage Scenarios} \label{sec:use_cases}

We envision two main usage scenarios for the use of \linebreak \gpdai, namely data exploration and software development. 
In both scenarios, the user may benefit from a different workflow. 
\gpdai\ was designed to provide a smooth user experience in both cases.

  \paragraph{Software development} When developing software, and in particular data pipelines, the programmer's main concern will be the generation of elementary transformations to be used on input DataFrames.
  While the user will also need to perform a \texttt{chat()} operation for each transformation, their workflow will be characterized by a heavy usage of the \texttt{inspect()} and \texttt{improve()} operations. 
  This workflow is facilitated by the global \texttt{safe\_mode} parameter. 
  In this case, the default behavior of \gpdai\ will be to print the generated code. 
  The programmer will therefore need to call the \texttt{execute()} function explicitly.
  Once the programmer is satisfied with the code generated, they can save it into a local function by using the \texttt{inject()} function. 
  
  \paragraph{Data exploration} During data exploration, the user can benefit from \gpdai's ability to plot or query the \linebreak DataFrames to obtain new insights on the data. 
  These operations are characterized by an intensive usage of the \texttt{chat()} and \texttt{execute()} functions.
  In a Jupyter notebook environment, composed of multiple cells, the caching mechanism presented in Section~\ref{sec:caching} will allow the user to execute or re-execute the cell in a random order, and still receive a deterministic result despite the stochastic nature of the LLM. 
  In order to facilitate the data exploration procedure, the default behavior of \gpdai\ is to perform automatically a call to \texttt{execute()} after every sequence of \texttt{chat()} and \texttt{improve()} which are in the same Python expression.

\section{Usecase Example} \label{sec:examples}

In this section, the usage of \gpdai\ is illustrated by adapting a \gpd\ tutorial from \textsc{hatarilabs}\footnote{\url{https://hatarilabs.com/ih-en/introduction-to-python-and-geopandas-for-flooded-area-analysis-tutorial}}.
This tutorial is about the analysis of flooded areas in the vicinity of the city of Boise. 
It involves plotting, joining and spatial analysis of multiple \gpd\ DataFrames.
The reader can run the examples provided hereunder as they are available on the project's GitHub repository in a Jupyter notebook.

\paragraph{\gdfai s initialization}
First, the user needs to load the three data sources into \gdfai s.
\begin{lstlisting}[style=mypython, basicstyle=\ttfamily\footnotesize]
floodedAreas = gai.GeoDataFrameAI(floodedAreas_gdf)
highways =  gai.GeoDataFrameAI(highways_gdf, description="This dataset contains information about roads.")
facilities = gai.GeoDataFrameAI(facilities_gdf)
\end{lstlisting}

\paragraph{Analytical queries on \gdfai s}
The user can perform some simple analytical queries. 
The results produced by \linebreak \gpdai\ are indicated in comments.
\begin{lstlisting}[style=mypython, basicstyle=\ttfamily\footnotesize]
floodedAreas.chat("What are the the bounds of the flooded areas.")
# [-116.5164, 43.6623, -116.2989, 43.6948]
facilities.chat("Find unique amenities.")
# ['school', 'hospital', 'fire_station']
\end{lstlisting}

\paragraph{Join queries and plot on \gdfai s}

All datasets can be plotted individually with the following queries:
\begin{lstlisting}[style=mypython, basicstyle=\ttfamily\footnotesize]
floodedAreas.chat("Plot the flooded areas.")
highways.chat("Plot the roads.").improve("Add a legend. Roads with the same name should have the same color.").improve("Improve the legend, it does not fit in its box. Make it scrollable.")
facilities.chat("Plot the facilities.", return_type=Figure)
\end{lstlisting}
Thanks to the corpus of geospatial programming examples, the \ai\ agent tends to plot the different \gdfai\ using specialized libraries such as Folium Maps. 
For illustration purposes, we specified in line 3 to produce a matplotlib Figure.
To produce a clearer map, line 2 needed further refinement using two \texttt{improve()} iterations. 
While it is also possible to obtain similar results using only one call to the \texttt{chat()} function by improving the query provided, this example showcases a real experiment, in which the user first performs an initial request, and then performs the desired refinement based on the intermediary results.
The \ai, however, successfully generated a map of the roads with a scrollable legend, as required.

Often, the user may need to perform operations based on multiple \gdfai\ objects. Two examples are provided hereunder.
\begin{lstlisting}[style=mypython, basicstyle=\ttfamily\footnotesize]
floodedFacilities = facilities.chat(
  "Add a Flooded column to the facilities based on whether they are in the flooded areas",
  floodedAreas,
)
floodedFacilities.chat("Export to the Out/floodedSchools.gpkg. Keep only the facilities flooded.", return_type=None)
\end{lstlisting}
In this example, the user creates a new \gdfai\ object based on the ``facilities'' but with an additional column based on the ``floodedAreas'' \gdfai.
He later uses this information to filter the \gdfai\ to keep only flooded facilities and to export the result in a GeoPackage file.

\begin{figure*}[t]
\centering
 \includegraphics[width=0.95\textwidth]{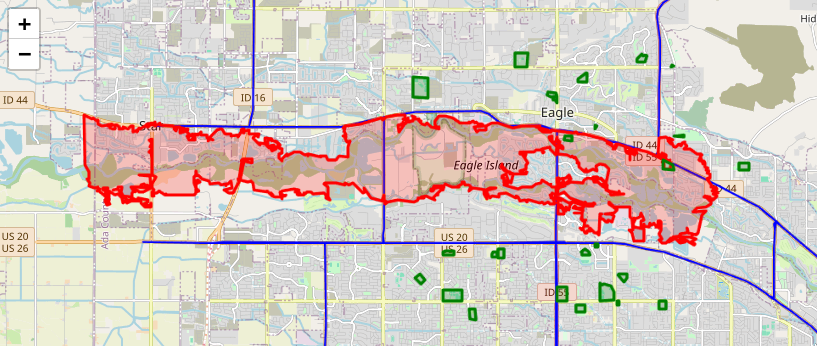}
 \caption{Folium Map of the three \gdfai s. Flooded areas, facilities and highways are displayed in red, green and blue, respectively.}
 \label{fig:join_plot}
\end{figure*}

\begin{lstlisting}[style=mypython, basicstyle=\ttfamily\footnotesize]
highways.chat(
  "Plot the highways the schools, and flooded areas.",
  floodedAreas, 
  facilities,
  return_type=Map).improve("Save in map.html file")
\end{lstlisting}
This example illustrates the generation and display of a map containing the data of all three \gdfai s.
This map is represented in Figure \ref{fig:join_plot}.
Further, the user could save this map in an ``.html'' file for later usage.

\paragraph{Inspecting and injecting code}
The user can inspect the history of generated code.
\begin{lstlisting}[style=mypython, basicstyle=\ttfamily\footnotesize]
facilities.inspect()
\end{lstlisting}
\begin{lstlisting}[basicstyle=\ttfamily\footnotesize]
Prompt 1: Add a Flooded column ...
Code 1:

import geopandas
import geopandas as gpd

def execute(df_1, df_2) -> GeoDataFrame:
    """Add a Flooded column to the ...

    :param df_1: GeoDataFrame containing facilities.
    :type df_1: geopandas.GeoDataFrame
    :param df_2: GeoDataFrame containing flooded areas.
    :type df_2: geopandas.GeoDataFrame
    :return: GeoDataFrame with a new 'Flooded' ...
    :rtype: geopandas.geodataframe.GeoDataFrame
    """
    df_1['Flooded'] = df_1.intersects(df_2.unary_union)
    return df_1
\end{lstlisting}

If the user is satisfied with the code and the result, they can materialize the operation in a function as follows:
\begin{lstlisting}[style=mypython, basicstyle=\ttfamily\footnotesize]
facilities.inject("flooded")
\end{lstlisting}
which will print the following instructions:
\begin{lstlisting}[basicstyle=\ttfamily\footnotesize]
Manual injection procedure...
First add, if not already present, the following import 
statement:
import ai
Then replace the following code with the function call:
ai.flooded(gdf1, gdf2, ...)
Make sure to adjust the function call with the correct 
parameters.
\end{lstlisting}

The function \texttt{flooded()} is now written locally and can be called again as follows:
\begin{lstlisting}[style=mypython, basicstyle=\ttfamily\footnotesize]
import ai

ai.flooded(facilities_gdf, floodedAreas_gdf)
\end{lstlisting}

\section{Conclusion} \label{sec:ccl}
This paper introduces the \gpdai\ library, offering an innovative approach for the exploitation of LLM assistants in geospatial software development and data analysis. By extending the GeoDataFrame with a conversational interface, a concept that we call \emph{smart class}, we demonstrated how the desired design principlaes of AI code copilots of context-awareness, low cognitive load, and privacy can be effectively addressed. Further, we also identified and tackled key technical challenges: supporting feedback loops through a stateful conversation model, achieving seamless integration into Python development environments, ensuring consistency across runs despite the stochastic nature of LLMs, and improving geospatial code generation through domain-specific augmentation techniques such as RAG and curated examples. 

We formaly defined an extension of the \gdf\ class with the support of a minimal set of functions enabling a user-friendly exploitation of an \ai. 
As part of our contributions, we provide an open-source implementation of \gpdai\ available on PyPI. While our focus has been on GeoPandas, the underlying design principles can serve as a template for developing similar smart classes in other domains, such as raster processing, audio analysis, etc. 

Future work includes enhancing the system's robustness through fine-tuned geospatial code models, expanding support for more complex geoprocessing tasks, and exploring deeper integration with existing geospatial toolchains and IDEs. 
Furthermore, an interesting further enhancement of \gpdai\ consists in integrating custom or existing artificial data generators, alleviating the burden for users with data sensitivity concerns. 
Ultimately, \gpdai\ represents a step toward more intuitive, collaborative, and domain-aware AI-assisted programming in the geospatial ecosystem.

\begin{acks}
The research leading to the results presented in this paper
has received funding from the European Union’s funded
projects MobiSpaces under Grant No. 101070279 and  EMERALDS under Grant No. 101093051.

This work was inspired by the open-source GitHub repository \textsc{Pandas-AI}\footnote{\url{https://github.com/sinaptik-ai/pandas-ai}} .  While the repository lacks a formal design and theoretical foundation, it motivated the idea of extending the DataFrame with a chat interface. In this work, we have systematically developed and formalized this concept.
\end{acks}

\bibliographystyle{ACM-Reference-Format}
\bibliography{bibli}


\end{document}